\documentclass[conference]{IEEEtran}
\IEEEoverridecommandlockouts
\usepackage{cite}
\usepackage{amsmath,amssymb,amsfonts}
\usepackage{algorithmic}
\usepackage{graphicx}
\usepackage{subcaption}
\usepackage{textcomp}
\usepackage{xcolor}

\usepackage[a4paper, total={184mm,239mm}]{geometry}
\def\BibTeX{{\rm B\kern-.05em{\sc i\kern-.025em b}\kern-.08em
    T\kern-.1667em\lower.7ex\hbox{E}\kern-.125emX}}

\newcommand*\circled[1]
            {\raisebox{.3pt}{\textcircled{\raisebox{-.9pt} {#1}}}}

\begin{document}
\title{\huge A Hybrid-Domain Floating-Point Compute-in-Memory Architecture for Efficient Acceleration of High-Precision Deep Neural Networks
}

\author{Zhiqiang Yi, Yiwen Liang, Weidong Cao \\Department of Electrical and Computer Engineering, The George Washington University, DC, US}

\maketitle

\begin{abstract}

Compute-in-memory (CIM) has shown significant potential in efficiently accelerating deep neural networks (DNNs) at the edge, particularly in speeding up quantized models for inference applications.
Recently, there has been growing interest in developing floating-point-based CIM macros to improve the accuracy of high-precision DNN models, including both inference and training tasks. 
Yet, current implementations rely primarily on digital methods, leading to substantial power consumption.
This paper introduces a hybrid domain CIM architecture that integrates analog and digital CIM within the same memory cell to efficiently accelerate high-precision DNNs.
Specifically, we develop area-efficient circuits and energy-efficient analog-to-digital conversion techniques to realize this architecture. 
Comprehensive circuit-level simulations reveal the notable energy efficiency and lossless accuracy of the proposed design on benchmarks.

\end{abstract}

\iffalse

\fi

\section{Introduction}
\label{sec:intro}

Compute-in-memory (CIM) holds significant promise for efficiently accelerating deep neural networks (DNNs) at the edge by bringing computation and memory closer together to reduce energy-intensive data movement that occurs in conventional von Neumann architecture\cite{wu2023floating, su20228, cao2023d, cao2021neural}.
Although numerous existing works have focused on building integer-based (INT-based) CIM macros\cite{dong202015,TSMC_CIM,SRAM_CIM_INT_1,sun2023efficient,9162917} to improve the energy efficiency of quantized DNN models, there is growing interest in developing floating-point-based (FP-based) CIM macros to enhance the accuracy of high-precision DNN models, including both inference and training applications.
Yet, most current FP CIM methods rely on digital implementation, leading to significant power consumption\cite{SRAM_CIM_FP_1,Sparse_intense,reCIM}.
Novel approaches are needed to bridge this gap, ensuring both high accuracy and energy efficiency for increasing high-precision DNN applications.

This paper proposes a novel hybrid-domain FP CIM architecture to remarkably improve energy efficiency while maintaining lossless accuracy for high-precision DNN applications.
The proposed architecture is based on a key observation that has been significantly overlooked in conventional designs of FP CIM macros.
It is observed that the FP arithmetic can be intrinsically divided into two parts: (1) the computation-intensive multiplication (sub-MUL), contributing less than $1/4$ to FP products, and (2) the computation-light addition (sub-ADD), contributing more than $3/4$ to FP products, as elaborated in Section~\ref{sec: addition}. 
We harness this insight to strategically integrate both analog CIM (for energy-efficient sub-MUL) and digital CIM (for accurate sub-ADD) on a unified hardware substrate.
This hybrid-domain CIM strategy, to the best of our knowledge, optimally combines the strengths of analog and digital CIM to achieve state-of-the-art energy efficiency by maintaining equivalent accuracy compared to fully digital baselines.
The key contributions of the work are listed below.

\begin{itemize}

\item We propose a novel hybrid-domain FP CIM architecture based on static random-access memory (SRAM) that integrates both digital and analog CIM within a unified macro to accelerate high-precision DNN applications. 
 
\item We develop detailed circuit schematics and physical layouts to implement this architecture. 
We optimize the local computing logic with minimal area overhead and minimize the energy cost of analog-to-digital conversion.

\item Experimental evaluations with comprehensive circuit-level simulations demonstrate the exceptional energy efficiency and accuracy of the proposed architecture compared to conventional fully digital baselines.
\end{itemize}

\section{Background and Related Work}
\label{sec:Related Work}

\subsection{Floating-Point Arithmetic Primitives}

FP arithmetic operations are crucial to high-precision DNNs. 
% models.

\begin{figure}[!t]
    \centering
    \includegraphics[width=0.3\textwidth]{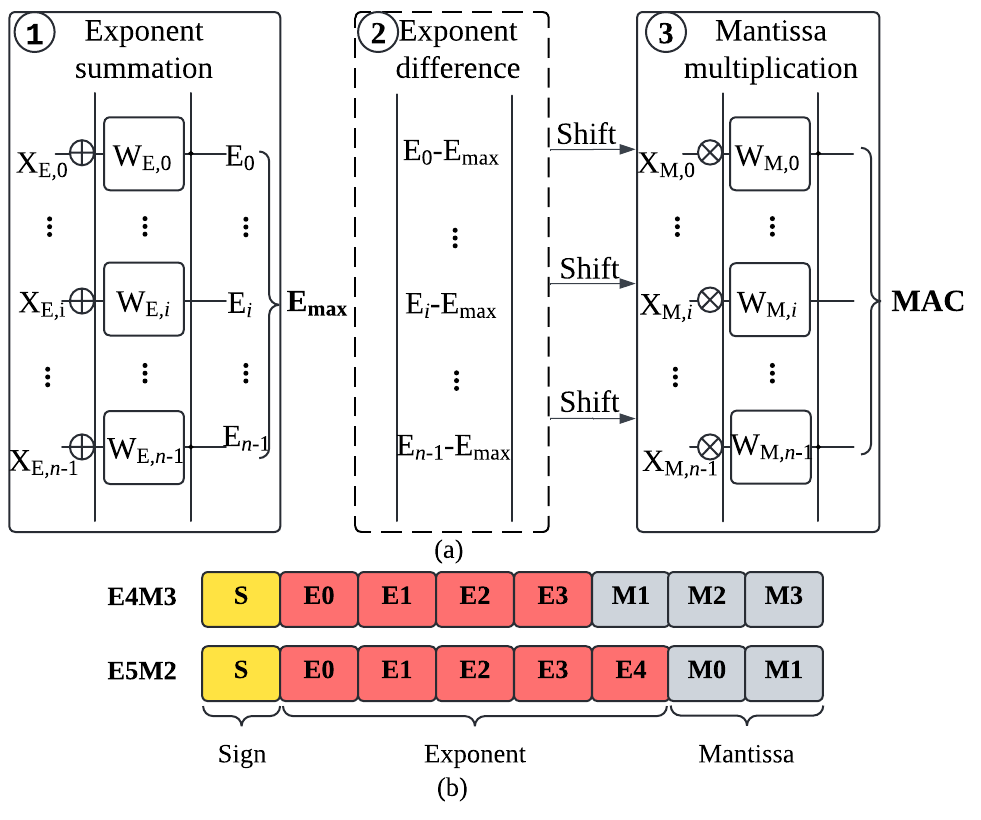} % 
     \vskip -12pt
    \caption{(a) Illustration of conventional FP SRAM CIM architecture. $X_{E,i}$ and $W_{E,i}$ are the exponent parts of activation and weight. $X_{M,i}$ and $W_{M,i}$ are the mantissa parts of activation and weight. \textcircled{1}, \textcircled{2}, and \textcircled{3} are circuit-level representations of the steps in Eq.~\eqref{eq: fp_add}, (b) FP8 format (E4M3 and E5M2).}
    \label{fig: FP_intro}
     \vskip -15pt
\end{figure}

\noindent{\textbf{FP format:}} A general FP number in scientific notation is expressed as $f=(-1)^{S}\cdot 2^{E}\cdot 1.M$,
where $S$ ($S=0$ or $S=1$), $E$, and $M$ ($M\in(0,1)$) represent the sign, exponent, and mantissa (fraction) of the number, respectively.
The `1' before $M$ is a hidden bit that is not explicitly shown in the binary format of an FP number. 
$E$ is the actual exponent, with an offset applied to the exponent encoded in the standard format.
For example, in the IEEE 8-bit FP format (FP8, E4M3 or E5M2), $S$ is 1-bit, $E$ is 4(5)-bit, $M$ is 3(2)-bit (Fig.~\ref{fig: FP_intro}(b)).

\noindent{\textbf{FP multiplication:}} The multiplication of FP numbers is a straightforward process, involving exponent addition (\circled{1}) and mantissa multiplication (\circled{2}). 
For simplicity, the example below demonstrates this using two positive FP numbers.
% \vskip -6pt
\begin{equation}
             (2^{E_0}\cdot1.M_0) \cdot (2^{E_1}\cdot1.M_1) = \underset{\text{\circled{1}}}{\underbrace{(2^{E_0 +E_1})}} \cdot \underset{\text{\circled{2}}}{\underbrace{(1.M_0\cdot 1.M_1)}}. 
    \label{eq: fp_mul}
\end{equation}
    \vskip -3pt
\noindent{\textbf{FP addition:}} However, the addition of floating-point (FP) numbers is more complex and can be expressed as follows:
% Eq.~\eqref{eq: fp_add} shows an example using two positive numbers for simplicity.
\begin{equation}
             2^{E_0}\cdot1.M_0+2^{E_1}\cdot1.M_1 = \underset{\text{\circled{1}}}{\underbrace{ 2^{{E_{\max}}}}} \cdot \underset{\text{\circled{3}}}{\underbrace{
\left\{ 
             \begin{array}{lr}
              \underset{\text{\circled{2}}}{\underbrace{2^{{E_{0}}-{E_{\max}}}}} \cdot 1.M_{0} &  \\
                +&  \\
              \underset{\text{\circled{2}}}{\underbrace{2^{{E_{1}}-{E_{\max}}}}} \cdot 1.M_{1}
             \end{array}
\right.}}.
    \label{eq: fp_add}
\end{equation}
The process begins with alignment, where the maximum exponent, $E_{\max}=\max\{E_0, E_1\}$,  is identified among all exponents (\circled{1}), followed by calculating the exponent difference, $E_{0(1)}-{E_{\max}}$, for each exponent (\circled{2}).
Next, the mantissa is shifted to the right based on this exponent difference and then summed (\circled{3}).
Finally, the result undergoes additional steps, such as truncation, to conform to the standard FP format.

\noindent{\textbf{FP vector-matrix multiplication:}}  By generalizing Eq.~\eqref{eq: fp_add} to the accumulation of $n$ FP numbers, where each number, $2^{E_i}\cdot 1.M_{i}$ is assumed to be a product of a weight-activation pair, i.e., $(-1)^{W_{S,i}}\cdot2^{W_{E,i}}\cdot 1.W_{M,i}$ and $(-1)^{X_{S,i}}\cdot 2^{X_{E,i}}\cdot 1.X_{M,i}$ based on Eq.~\eqref{eq: fp_mul} (i.e., $E_i=W_{E,i}+X_{E,i}$ and $1.M_{i}=1.W_{M,i}\cdot1.X_{M,i}$), the vector-matrix multiplication (VMM) of FP numbers is achieved. 
This operation is fundamental to FP DNN models.
FP enables high-precision computations in DNN models, supporting the highest accuracy and best training quality~\cite{micikevicius2022fp8}.
Thus, efficient hardware acceleration for FP VMM is highly desirable.

\iffalse

\subsection{Computing-in-Memory for FP Operation}
Figure 1 illustrated how digital CIM accelerate the FP VMM. 

Recent work are interested in using digtial CIM structure to accelerate the FP DNN models.  Tu et al. 's CIM macro operates all FP computation, but they transform the Float-point to Fixed-point first.  Lee et al. also implement a CIM macro that it only computes part of FP and the rest is operated in digital circuits. Digital computing provides exceptional accuracy for FP operation. However, digital FP CIM macros have been limited to a few edge applications due to the power consumption. 
Circuit-level techniques, 
Although there are many different optimization research, they primarily focus on the accuracy of computation and accelerate every exponent sum and mantissa multiplication in the digital domain, resulting in remarkable, yet necessary energy waste.

Our work based on the weight theory, the maximum error for mantissa multiplication can not exceed 1/4, so we take a more energy friendly, but precision loss way -- analog circuit to accelerate the multiplication. As for mantissa addition, which contribute 3/4 to the result, we keep operating it in digital domain.

\fi

\subsection{Compute-in-Memory for FP DNNs}

Previous work has explored CIM acceleration for FP DNN models using emerging non-volatile memory (NVM) devices~\cite{luo2020accelerating, PF_CIM_RRAM_0}.
More recently, static random-access memory (SRAM) has become a cornerstone in the design of CIM systems for practical applications due to its exceptional accuracy in digital computing, along with superior performance, power efficiency, and area benefits~\cite{reCIM, Sparse_intense, SRAM_CIM_INT_1, SRAM_CIM_FP_1}.
For instance, a previous study~\cite{reCIM} introduced a reconfigurable FP/INT CIM processor capable of supporting both BFloat16 (BF16)/FP32 and INT8/16 in the same digital CIM macro.
Another study~\cite{Sparse_intense} proposed dividing FP operations into high-efficiency intensive-CIM and flexible sparse-digital parts, based on the observation that most FP exponents cluster within a narrow range. 
Circuit-level techniques such as time-domain exponent summation mechanisms have also been employed to enhance the energy efficiency of CIM macros~\cite{SRAM_CIM_FP_1}.

Nonetheless, these prior works share the common simplified architecture to accelerate FP VMM, as illustrated in Fig.~\ref{fig: FP_intro}(a), where all operations are performed in the digital domain.
The process begins with the alignment of exponents by summing the exponent parts of the weight-activation pairs (${W_{E,i}}$ and ${X_{E,i}}$, Step \circled{1}) and calculating the exponent difference between each sum, $E_i$,  and the maximum exponent, $E_{\max}$ (Step \circled{2}).
These exponent differences ($E_i-E_{\max}$) are then used to shift the mantissa parts of the activations, which are subsequently multiplied and accumulated (Step \circled{3}).
As shown in Fig.~\ref{fig: FP_intro}(a), these steps align with the operations described in Eq.~\eqref{eq: fp_add}.
Yet, our analysis reveals that by leveraging the inherent properties of FP arithmetic and the resilience of DNN models to minor computation errors, there is substantial untapped potential to use approximate computing (e.g., analog computing) to accelerate FP DNNs without sacrificing accuracy (Section~\ref{sec: addition}).

% This insight presents valuable opportunities to further enhance the performance of FP SRAM CIM macros.

\begin{figure*}
    \centering
    \includegraphics[width=0.8\textwidth]{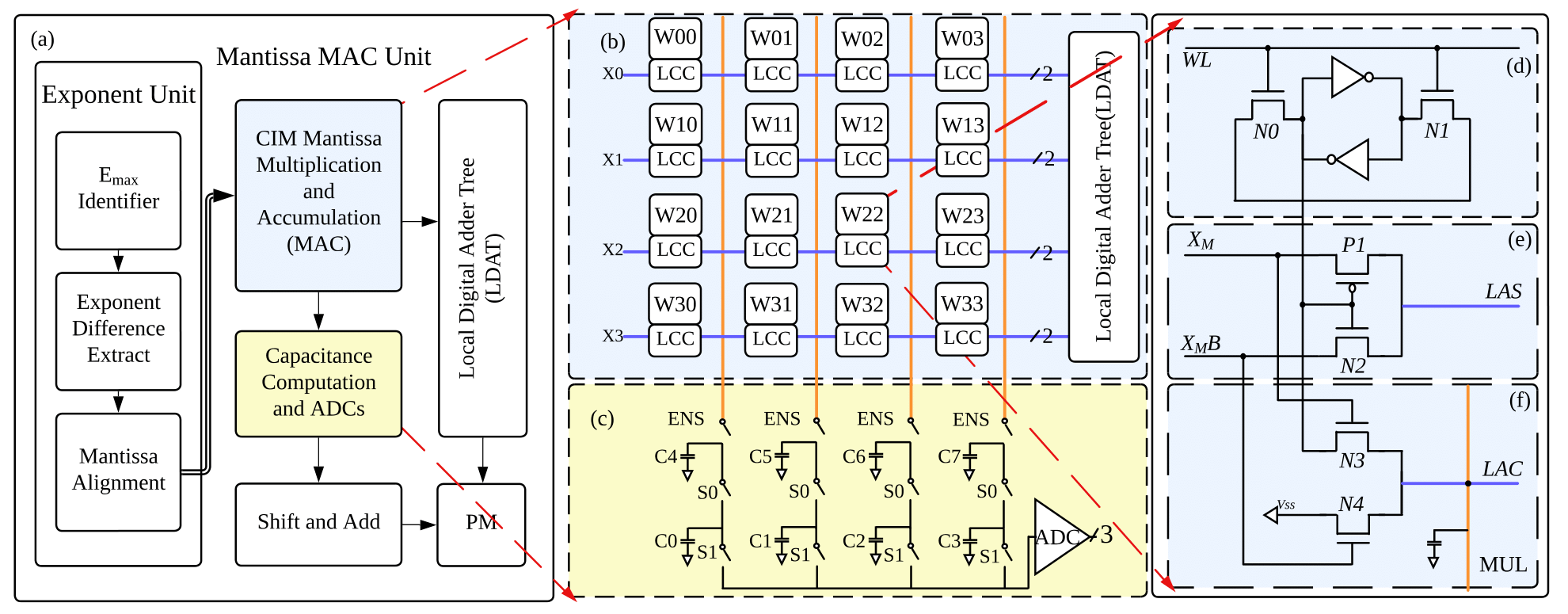} % 
     \vskip -6pt
    \caption{(a) Architecture overview of the proposed hybrid-domain FP CIM macro, (b) Hybrid-domain FP SRAM CIM mantissa unit, (c) Switched-capacitor array with the Flash ADC, (d) 6T SRAM, (e) pseudo XOR gate, (f) pseudo AND gate, LAC for sub-add (blue line) and MUL for ACIM (orange line).}
    \label{fig: Architecture}
     \vskip -15pt
\end{figure*}

\section{Hybrid-Domain FP CIM Architecture Design}

\subsection{Insight: Addition is More Important}
\label{sec: addition}

Through a careful examination of FP arithmetic, we have identified a lightweight approach for implementing SRAM CIM macros in a hardware-friendly manner, enhancing energy efficiency while preserving the accuracy of FP DNNs. To illustrate this key concept, we decompose the conventional FP mantissa multiplication in Eq.~\eqref{eq: fp_mul} into two parts as follows:
\begin{equation}
\begin{split}
&(1.M_0\cdot 1.M_1)=(1+M_0)\cdot (1+M_1)\\
 &=(\underset{\text{sub-ADD}}{\underbrace{1+M_0+M_1}}+ \underset{\text{sub-MUL}}{\underbrace{M_0 \cdot M_1}}).
\end{split}
    \label{eq: fp_mul_1}
\end{equation}
Here, $(1+M_0+M_1)$ and $(M_0\cdot M_1)$ are defined as mantissa sub-addition (sub-ADD) and mantissa sub-multiplication (sub-MUL), respectively.
Based on this decomposition and given that $M_{0(1)}\in(0,1)$, we analyze the significance of sub-MUL in mantissa multiplication\cite{cao2024addition}, i.e.,
\begin{equation}
\dfrac{M_0 \cdot M_1}{(1+M_0)\cdot(1+M_1)}=\dfrac{1}{(1+1/M_0)\cdot(1+1/M_1)}\leq \dfrac{1}{4}.
\label{eq: fp_mul_sig}
\end{equation}
Here, the `$=$' holds true if and only if $M_0=M_1	\rightarrow 1$.
Eq.~\eqref{eq: fp_mul_sig} shows that for a single weight-activation pair, if the accuracy of sub-ADD is maintained, the \textit{total computation error of mantissa multiplication (and thus the FP product) will not exceed $1/4$ of the true value, even with the aggressive removal of sub-MUL operations}. 
Our analysis reveals that while mantissa sub-MUL operations are computationally intensive, they often affect only a minority of FP products. 
In contrast, mantissa sub-ADD, though computationally lighter, constitutes the majority of FP products. Addition is also far more energy-efficient than multiplication; for instance, INT8 addition consumes only about 10\% of the energy required for INT8 multiplication, as shown in previous work~\cite{TPU}.
This key insight--`Addition is More Important'--guides our design strategy for FP SRAM CIM macro: \textbf{dedicating digital resources to ensure the precision of compute-light mantissa sub-ADD operations while leveraging energy-efficient analog computing for the more compute-intensive mantissa sub-MUL operations}.

\label{sec:method}

\subsection{Overall Macro Architecture}

With the above key insight, we integrate analog CIM into existing well-established digital FP CIM architectures~\cite{SRAM_CIM_FP_1,SRAM_CIM_INT_1} based on SRAM, achieving significant improvements in energy efficiency with minimal area overhead.
Fig.~\ref{fig: Architecture} illustrates the proposed architecture, designed for FP8 DNN models (E4M3) as an example.
Note that FP8 demonstrates state-of-the-art energy efficiency in DNN applications while preserving accuracy\cite{micikevicius2022fp8}. 
The exponent unit comprises core components such as the exponent summation array, $E_{\text{max}}$ identifier, exponent difference extractor, and mantissa alignment, in line with previous designs~\cite{time_domain,SRAM_CIM_FP_1}. 
The details are thus omitted here.
We focus on the mantissa part.
We introduce a novel hybrid-domain MAC unit to precisely manage mantissa sub-ADD in the digital realm while enhancing the energy efficiency of mantissa sub-MUL through analog computing.

Specifically, we demonstrate a hybrid-domain CIM mantissa MAC array with a $4 \times 4$ size as a proof of concept (Fig.~\ref{fig: Architecture}(b)). 
This $4 \times 4$ array is ideally suited for a 4-bit weight mantissa in E4M3 FP8 formate, including a 1-bit sign indicator.
Each bit of a weight mantissa $W_{M,i}$ is stored within the same row but across different columns of the 6T-SRAM cell array, ordered from the most significant bit (MSB) to the least significant bit (LSB).
There are two key design innovations in this MAC array: (1) an area-efficient hybrid-domain local computing cell (LCC) is embedded into each memory cell to facilitate both sub-ADD and sub-MUL;
(2) the analog partial sums of sub-MUL are accumulated into the analog domain to minimize analog-to-digital conversions.
In the following subsections, we will introduce how sub-ADD and sub-MUL are enabled with these circuit-level design innovations.

\subsection{Digital CIM -- Sub-ADD}

The sub-ADD operation is achieved through two pseudo-digital logic gates (a pseudo AND gate and a pseudo XOR gate) in the LCL. 
These gates, each composed of only two transistors to minimize area overhead, form a half-adder. 
Additionally, the pseudo AND gate is reused in a time-multiplexed manner for the sub-MUL operation, as shown in the next section. 
Using a bit-serial input, a 2-bit partial sum of the activation and weight mantissas is generated along each row in the $j^{\text{th}}$ cycle. When $X_{m,i}[j] = 1$, $\text{LAC}_{i,j} = W_{M,i}[j] - V_{\text{th}}$; when $X_{M,i}[j] = 0$, $\text{LAC}_{i,j} = V_{\text{SS}}$. 
Here, LAC represents the carry bit in the summation of the activation and weight mantissa, specifically for the $j^{\text{th}}$ bit of the mantissa part.
Similiarly, pseudo XOR is equivalent to normal XOR gate in half adder.
The 2-bit partial sum will be accumulated with the local adder to obtain the full sum of sub-ADD.
%Additionally, pseudo AND is a time-multiplexing logic gate. 

\subsection{Analog CIM -- Sub-MUL}

Fig.{~\ref{fig: Architecture}}(b) illustrates the sub-MUL operation, which is divided into two key steps: first, the pseudo AND gate in each cell performs the multiplication between the activation mantissa and weight mantissa; second, the resulting multiplication values from each cell are summed along each bit line.

% \textcolor{red}{The logic here be: (1) briefly introduce how the pseudo AND gate in each cell performs the multiplication; (2) how analog partial sums of each BL can be accumulated with capacitor; (3) How to compensate for the mismatch of the capacitor.}

\subsubsection{Analog multiplication by reusing the AND gate}
The pseudo AND logic gate obtains and sends the mantissa multiplication result to the capacitor computation unit through orange line on Fig.~\ref{fig: Architecture}(f). 
The analog calculation relies on Kirchhoff’s law: voltage on each GBLB decreases due to the discharge current from cells and increases with the charging current. The final voltage on each bit line reflects the net current of charged and discharged cells on each GBLB.

\subsubsection{Analog accumulation of partial sums}

We use the switched-capacitor array to efficiently accumulate analog partial sums of sub-ADD across bitlines.
%The capacitor charge-sharing computation integrates the analog multiplication signals, enabling efficient accumulation. 
This mechanism reduces the required analog-to-digital conversions from four to one per input cycle, significantly enhancing energy efficiency.
Assume the voltage of each $\text{GBLB}[i]$ is $V_i$ and the capacitance is $C_i$, then we have $Q_i= V_i \cdot C_i$.
% \begin{equation}
% Q_i= V_i \cdot C_i
% \end{equation}
All charges are collected first in the computation capacitance region as $Q_{total} =\sum_{i=0}^{3} V_i \cdot C_i$.
% \begin{equation}
% Q_{total} =\sum_{i=0}^{3} V_i \cdot C_i.
% \end{equation}
When switch S1 is activated, charge sharing occurs at this stage, resulting in $V_{o} =({\sum_{i=0}^{3} V_i \cdot C_i})/({\sum_{i=0}^{3} \cdot C_i})$.
% \begin{equation}
% V_{o} =\frac{\sum_{i=0}^{3} V_i \cdot C_i}{\sum_{i=0}^{3} \cdot C_i}.
% \end{equation}
Hence, $V_o$  represents the 1-bit multiplication of the activation mantissa and the 4-bit weight mantissa.
\iffalse
The final result is given by the formula:
\begin{equation}
\sum_{n=0}^{3}V_{o,n}=\sum_{n=0}^{3} \sum_{x=0}^{3} 2^m \cdot X_n[m] \cdot M_n.
\end{equation}
\fi

Fig.~\ref{fig: Architecture}(c) illustrates the multi-bit weight realization using charge sharing among computational capacitors. 
From the LSB to the MSB, the GBLB (orange line) connects to computation capacitors $C_0, C_1, C_2, C_3$ in a 1:2:4:8 capacitance ratio. 
Initially, all capacitors are pre-charged to ${V}_{\text{DD}}$.
To maintain a total of 8 units of capacitance on each column, the compensation capacitors $C_4, C_5, C_6, C_7$ are configured with ratios of 7:6:4:0, respectively.
After processing by the computation capacitors, the final voltage represents the multiplication of the 4-bit input with 1-bit weights. 
This voltage, representing a $4\times4$-bit multiply-accumulate (MAC) result, is isolated from the GBLB to prevent signal interference and converted to a 3-bit digital output by the Flash ADC.

The Flash ADC utilizes area-efficient sense amplifiers (SAs) instead of traditional analog comparators to reduce both area and energy consumption. Each 3-bit Flash ADC is composed of 7 SAs.
In a $4\times4$ CIM array, the multiplication result voltage on each global bit line buffer (GBLB) may yield 4 possible values, resulting in $4\times16$ potential outputs, which would typically require a 6-bit ADC for full encoding.
To optimize efficiency, we encode only the 3 most significant bits (MSBs). 
This selective encoding not only minimizes encoding area overlap due to nonlinearity but also reduces the impact on accuracy, ensuring effective performance with lower resource demands.
{Our simulation results in Section~\ref{sec:Results} show that 3-bit ADC is sufficient to achieve loss-less accuracy for inference/training.

\section{Evaluation}
\label{sec:Results}
\subsection{Computation Accuracy}

Since the exponent array is implemented in the digital domain without computation errors, we focus on presenting the computing accuracy of our mantissa MAC array.
We compare the results from circuit-level simulations through the CIM macro and its ground truth which
%based on a Python script.
% The ground truth of the test 
is obtained by simulating the computation flow through a Python script. 
We obtain the product of $4\times4$-b inputs and the $4\times4$-b weights.  
The computation error across multiple combinations of 4-b inputs and 4-b weights are shown in Fig.~\ref{computation error}.
It is small within 1.51\% range.
The result shows that the proposed hybrid-domain computing introduces minimal computation error.

\begin{figure}
    \centering
    \includegraphics[width=0.3\textwidth]{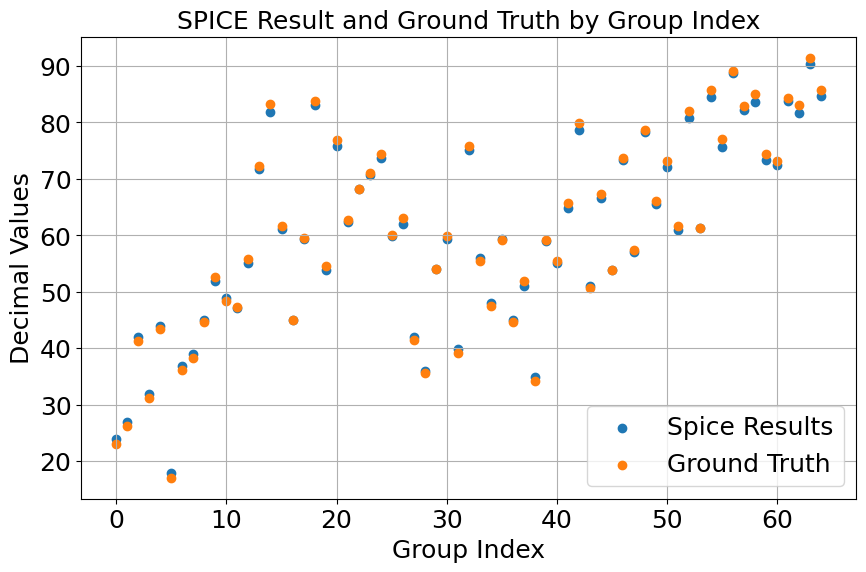} % 
     \vskip -6pt
    \caption{The
computation error across multiple combinations of 4-b inputs
and 4-b weight}
    \label{computation error}
     \vskip -15pt
\end{figure}
% This accuracy is mainly focus on charge accumulator and ADC encoder.

% \subsubsection{Baseline}
% The ground truth of test is obtained by simulating the computation flow through python script. We obtain the sum and product of a input $4\times4$ matrix X and a weight $4\times4$  matrix W, respectively.  

% \subsubsection{Result}

We then apply the computed error to FP products based on FP arithmetic and inject this error into FP8 models to examine its effect on model accuracy.
We evaluate accuracy across various FP DNN models using the MLPerf Edge Inference Benchmark Suite v4.0, which encompasses a diverse range of deep learning tasks. 
As shown in Table~\ref{tab:model_performance}, our proposed hybrid-domain architecture effectively maintains inference accuracy, with only approximately 1\% variation observed for ResNet and BERT models, and virtually no accuracy loss for RetinaNet. 
This outcome underscores our macro’s reliability in preserving system-level inference accuracy, likely due to the inherent robustness of these models.

% Specifically, we evaluate ResNet-50 v1.5 on ImageNet, RetinaNet on OpenImages, and BERT-large on SQuAD v1.1 undering FP16 format. 

\iffalse
We present the accuracy of our mantissa MAC  is to compare the multiplication of FP through CIM macro and its truth value. This accuracy is mainly focus on charge accumulator and ADC encoder.
\subsubsection{Baseline}
The ground truth of test is obtained by simulating the computation flow through python script. We obtain the sum and product of a input $4\times4$ matrix X and a weight $4\times4$  matrix W, respectively.  
%input Xm and weight Wm are 4x4 matrix, so there are 2^16 times 2^16 situations. we can randomly pick some of them for verification.
\subsubsection{Result}
To evaluate the system-level inference performance of our proposed SRAM CIM structure, we assess its accuracy across various FP DNN models using the MLPerf Edge Inference Benchmark Suite v4.0, which spans diverse deep learning tasks []. Specifically, we evaluate ResNet-50 v1.5 on ImageNet, RetinaNet on OpenImages, and BERT-large on SQuAD v1.1 undering FP16 format. Table  \label{model_performance}  shows that our HyCIM structure maintains inference accuracy effectively, with approximately a 1\% accuracy variation for ResNet and BERT, and nearly no accuracy drop for RetinaNet. This demonstrates that our macro reliably preserves system-level inference accuracy. This observation may be attributed to the inherent robustness of the models.

\fi

\begin{table}[h]
    \centering
        \caption{Model Performance on Different Datasets.} 
    \renewcommand{\arraystretch}{1.2} % 调整行距
             \vskip -6pt
    \begin{tabular}{|p{1.5cm}|p{1.5cm}|p{2.5cm}|p{1cm}|}
        \hline
        \textbf{Model} & \textbf{Dataset} & \textbf{Baseline} \textbf{Accuracy}& \textbf{Result} \\
        
        \hline
        ResNet50 & ImageNet & 76.01\% (top1 Acc)& 75.68\% \\
        \hline
        BERT-base & SQuAD v1.1 & 90.73\% (f1\_score)& 89.49\% \\
        \hline
        Retinanet & Open Images & 0.3759 (mAP)& 0.3709\\
        \hline
    \end{tabular}   \label{tab:model_performance}
             \vskip -12pt
\end{table}

\subsection{Energy Comparison}

We designed the proposed Mantissa MAC unit using 28 nm CMOS technology and compared its energy efficiency to a fully digital baseline, which includes a digital multiplier at each memory cell (enabled by a NAND gate), a local digital adder tree, an accumulator, and registers.
Energy data for the digital baseline is derived from the quantitative model in ~\cite{sun2023analog}. 
Table~\ref{tab:energy_breakdown} illustrates the comparison, showing our design achieves $1.53\times$ energy efficiency compared to the digital CIM.

% scale the sub-ADD and sub-MUL with capacitance, delay, area of single NAND2 gate. 

% The traditional fully digital CIM macro consists of a 6T-SRAM and a NAND gate.
% It totally has 10 transistors (4 PMOS and 6 NMOS) which is the same as our macro. 
% Thus, the configuration like capacitance, area and delay will be close. 
% The result of this comparison is convincing. 
% From the Table.\textcolor{blue}{ \label{energy_breakdown}} , it captured the energy of sub-ADD, sub-MUL and baseline which is DCIM.
% The energy efficiency of our design is 34.8\% than DCIM.

\iffalse

Energy breakdown from the Quantitative model of paper ~\cite{sun2023analog}, scale the sub-ADD and sub-MUL with capacitance, delay, area of single NAND2 gate. The traditional fully digital CIM macro consists of a 6T-SRAM and a NAND gate. It totally has 10 transistors (4 PMOS and 6 NMOS) which is the same as our macro. Thus, the configuration like capacitance, area and delay will be close. The result of this comparison is convincing. From the Table.{ \label{energy_breakdown}} , it captured the energy of sub-ADD, sub-MUL and baseline which is DCIM. The energy efficiency of our design is 34.8\% than DCIM.

\fi

\begin{table}[h]
    \centering
    \renewcommand{\arraystretch}{1.2} % 调整行距
    \caption{Energy Breakdown Estimation (fJ/MAC).}
                 \vskip -6pt
    \begin{tabular}{|p{1.2cm}|p{1cm}|p{1cm}|p{1.5cm}|p{1.1cm}|p{0.8cm}|}
        \hline
        \textbf{Operation} & \textbf{Add/} 
        \textbf{Multi}& \textbf{Adder Tree} & \textbf{Accumulator}/ \textbf{ADC}& \textbf{Register} & \textbf{Total} \\
        \hline
        sub-ADD  & 0.896 & 10.752 & 10.752 & 6.720 & 29.120\\
        \hline
        sub-MUL  & 1.792 & 8.064& 7.04& 5.376& 22.272 \\
        \hline
        Baseline  & 3.584 & 34.944 & 26.880 & 13.440  & 78.848\\
        \hline
        Efficiency & - & - & - &- & 1.53$\times$ \\
        \hline
    \end{tabular}  \label{tab:energy_breakdown}
             \vskip -12pt
\end{table}

\subsection{Area Breakdown Estimation}
%Area overhead

We finally show the area efficiency of the proposed mantissa MAC unit.
Each cell consists of a 6T-SRAM cell, a pseudo AND, and a pseudo XOR, totaling 11 transistors (4 PMOS and 7 NMOS). 
The SRAM cell and pseudo AND are symmetrical, resulting in a compact and efficient mantissa computing unit, as shown in Fig.~\ref{fig:mantissa_layouts}(a).
It leads to a 42.8\% area overhead of the initial 6T SRAM cell.
When expanded to configurations of two or four cells, the layout achieves even greater symmetry and space efficiency, as shown in Fig.~\ref{fig:mantissa_layouts}(b).
As noted, a traditional digital CIM macro, which uses NAND gates attached to each memory cell, employs a nearly identical number of transistors, thus resulting in minimal area variation between the two macros.

% As is mentioned above, a traditional DCIM macro has the same transistors number as our macro, only one transistor difference in doping type. Thus, there is no large area variation on these two macros.

\iffalse
The element mantissa MAC unit includes a 6T-SRAM cell, a pseudo AND and a pseudo XOR. It has totally 11 transistors(4 PMOS and 7 NMOS). Notice that the SRAM cell and pseudo AND are symmetric. 
Thus, a nearly high compact mantissa computing unit is constructed as Fig.~\ref{fig:mantissa_layouts}(a) is shown. The initial SRAM cell is surrounded by yellow square, so the area overhead of mantissa MAC is 42.8\% than the initial SRAM cell.
Continue to expand the structure to 2 cells, 4 cells, it can be placed in a more symmetrical structure as shown in ~\ref{fig:mantissa_layouts}(b),  thus saving more space.
As is mentioned above, a traditional DCIM macro has the same transistors number as our macro, only one transistor difference in doping type. Thus, there is no large area variation on these two macros.

\fi

% SRAM cell area: 635.04
% MAC cell area  : 906.84

\begin{figure}[ht]
    \centering
    \begin{subfigure}[b]{0.20\textwidth}
        \centering
        \includegraphics[width=0.8\linewidth]{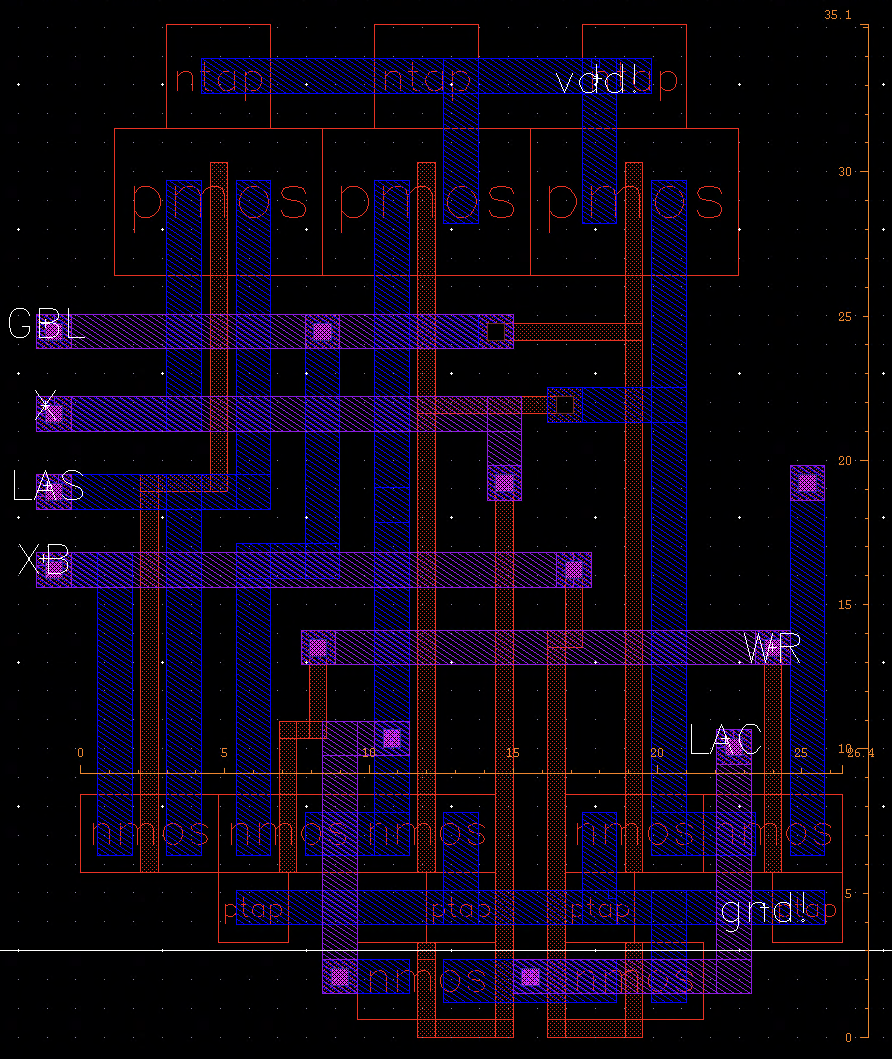} 
        \label{fig:mantissa_unit}
    \end{subfigure}%
    \hfill
    \begin{subfigure}[b]{0.215\textwidth}
        \centering
        \includegraphics[width=0.8\linewidth]{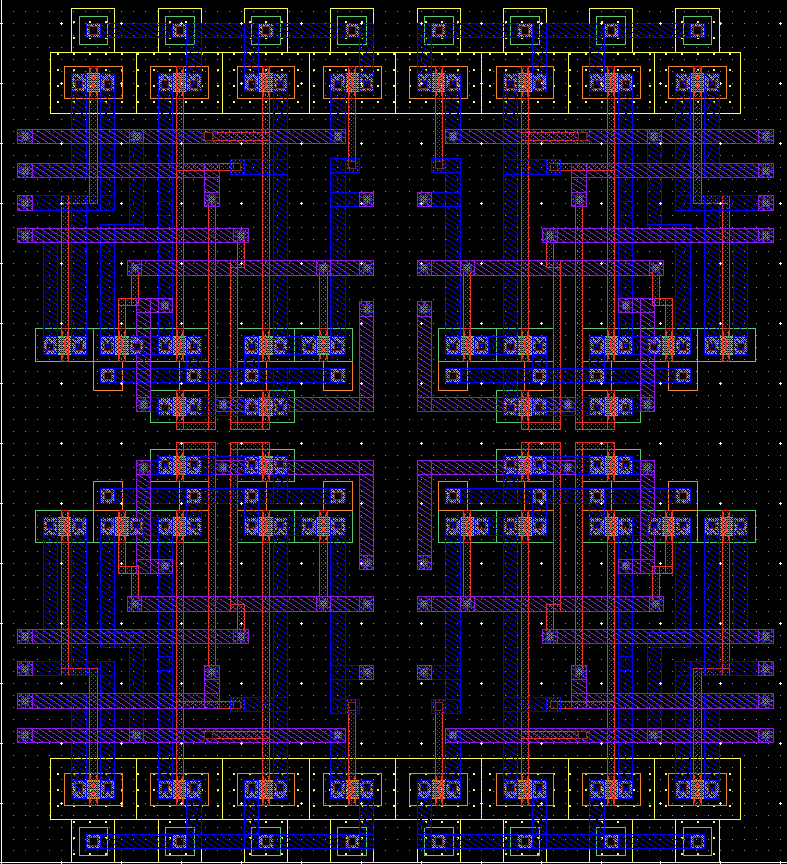} 
        \label{fig:four_cells}
    \end{subfigure}
         \vskip -6pt
    \caption{Mantissa MAC unit layouts: (a) Single mantissa computing unit, (b) Layout of 4 cells.}
         \vskip -15pt
    \label{fig:mantissa_layouts}
\end{figure}
 \section{Conclusion}

We propose a novel hybrid-domain FP CIM architecture based on SRAM, integrating both digital and analog CIM elements. 
Detailed circuit designs were developed for the hybrid-domain mantissa MAC unit, and circuit-level evaluations demonstrate notable energy efficiency gains compared to fully digital baselines, with minimal area overhead and negligible accuracy loss.

\iffalse

We propose a novel hybrid-domain FP CIM architecture based SRAM that integrates both digital and analog CIM. Since mantissa MAC is most complicated and energy-cost part in whole architecture, we developed some optimal structures to save energy and area and also we built a 4x4 array circuit schematics and physical layouts for verification. According to the rule that mantissa addition take 3/4 weight in FP multiplication, so we use fully digital circuit to ensure the accuracy, and a sub-add for adder structure which only increase 55.7\% area overhead; mantissa multiplication take only 1/4 weight in FP operation, so we use an analog circuit to lower the energy consumption and area overhead. Also we use a capacitance sharing way to decrease the operation times of shift and add so that we can obtain higher energy efficiency. Last, comprehensive evaluation show it improve energy efficiency and area overhead with accuracy.

\fi

%\bibliographystyle{IEEEtran}
%\bibliography{biography}

\end{document}